\documentclass[aps,prb,twocolumn, showpacs, superscriptaddress]{revtex4}
\usepackage{graphics}
\newcommand{\insertpic}[1]{\scalebox{0.31}{\rotatebox{-90}{\includegraphics{#1}}}}
\newcommand{\bvec}[1]{{\mathbf #1}}

\begin{document}
\draft 
\widetext

\title{Exchange anisotropy, disorder and frustration in diluted, predominantly ferromagnetic, Heisenberg spin systems}

\author{Chenggang Zhou}
\affiliation{Despartment of Electrical Engineering, Princeton University, Princeton, New Jersey 08544, USA}
\author{Malcolm P. Kennett}
\affiliation{TCM Group, Cavendish Laboratories, Cambridge University, Madingley Rd, Cambridge, CB3 0HE, UK } 
\author{Xin Wan}
\affiliation{National High Magnetic Field Laboratory, Florida State University, Tallahassee, Florida 32310, USA }
\altaffiliation[ Present address:] { Institut f\"ur Nanotechnologie, 
Forschungszentrum Karlsruhe, D-76021 Karlsruhe, Germany.}
\author{Mona Berciu}
\affiliation{Department of Physics and Astronomy, University of British Columbia, Vancouver, BC V6T 1Z1, Canada}
\author{R. N. Bhatt}
\affiliation{Department of Electrical Engineering, Princeton University, Princeton, New Jersey 08544, USA}

\date{\today}

\begin{abstract}
Motivated by the recent suggestion of anisotropic effective exchange
interactions between Mn spins in Ga$_{1-x}$Mn$_x$As (arising as a
result of spin-orbit coupling), we study their effects in diluted Heisenberg spin systems.  
We perform Monte Carlo simulations on several phenomenological model spin Hamiltonians, and investigate the extent to
which frustration induced by anisotropic exchanges can reduce the low
temperature magnetization in these models and the interplay of this
effect with disorder in the exchange. In a model with 
low coordination number and purely ferromagnetic (FM) exchanges, we find that the low
temperature magnetization is gradually reduced as exchange anisotropy is turned
on.  However, as the 
connectivity of the model is increased, the effect of small-to-moderate
anisotropy is suppressed, and the magnetization regains
its maximum saturation value at low temperatures unless the distribution 
of exchanges is very wide. To obtain significant suppression of the low temperature magnetization in a 
model with high connectivity, as is found for long-range interactions, 
we find it necessary to have both ferromagnetic and antiferromagnetic 
(AFM) exchanges (e.g. as in the RKKY interaction). 
This implies that disorder in the sign of the exchange interaction is much 
more effective in suppressing magnetization at low temperatures than exchange anisotropy.

\end{abstract}


\maketitle

\narrowtext
\section{Introduction}
\label{sec:intro}
It has recently been suggested that frustration effects may be
important for the magnetic properties of diluted magnetic
semiconductors (DMSs) such as
Ga$_{1-x}$Mn$_x$As.\cite{Zarand,Schliemann,Potashnik} 
The reported Curie temperatures ($T_c$) in these compounds continue to rise,
with a maximum of 140K recently reported for Ga$_{1-x}$Mn$_x$As,\cite{Edmonds}
with even higher temperatures reported for related materials.\cite{Hebard}
Aside from increasing $T_c$, 
it will also be important to have a thorough understanding of the different
aspects of their magnetic properties over a wide temperature range in order to
be able to construct optimized spintronic devices.
In theoretical analyses,
Zarand and Janko\cite{Zarand} showed that within the RKKY
approximation, a proper treatment of the spin-orbit coupling leads to
anisotropic exchanges between Mn spins.  Using this interaction they
found that the saturation magnetization is reduced by up to 50 \% at
low temperatures.  Experimentally it has been observed that in many
DMS, the saturation magnetization at low temperatures is not as large
as would be expected if all Mn moments were aligned (i.e. the
full saturation value).\cite{Oiwa} While it is likely that
magnetically inactive Mn, such as Mn interstitials,\cite{Yu} may
account for some of the suppression of the low-temperature
magnetization, the results of Zarand and Janko suggest that
anisotropic spin interactions may also play a significant role in
accounting for it. We carefully investigate this possibility in this
paper.

In a system consisting of localized spins (local moments) coupled to 
non-interacting fermions (carriers), where the spin-carrier interaction is a 
perturbation on the fermion Hamiltonian, the low-lying spin excitations can 
be described in terms of an RKKY interaction between the spins. In the case 
of DMS, the carrier density is low (in fact, lower than the density of 
local moments by a considerable factor, due to carrier compensation), 
and the carrier-spin interaction is quite strong (necessary to 
enable a high $T_c$). Consequently, the RKKY approximation 
appears unlikely to be appropriate for a {\it quantitative}
description of the ferromagnetism in Ga$_{1-x}$Mn$_x$As since the Fermi energy
is not necessarily much larger than the magnetic coupling. However,
the qualitative prediction of anisotropic exchange may be present
in more precise treatments.\cite{Zarand2}

In what follows, therefore, we assume that the magnetic properties of 
the system can be modeled in terms of an effective spin Hamiltonian, 
with a form that maintains the symmetry properties of the RKKY 
interactions obtained in the weak-coupling limit. 
Rather than calculate the anisotropy in the Mn-Mn effective exchange
within a microscopic model, we use a different approach to investigate
the relevance of anisotropy for systems of diluted spins.  We consider
phenomenological models with disordered, anisotropic exchanges between classical
Heisenberg spins placed randomly at low densities on a fcc lattice (corresponding to
the Ga fcc sublattice in (Ga,Mn)As), and study the
magnetic properties for several functional forms of the exchange
interactions.  In each case we consider various values of the disorder
and anisotropy.  We focus on whether full saturation in magnetization
is reached at low temperatures and also investigate the 
magnetic susceptibility, since this is known to be a
good experimental indicator of spin freezing.\cite{Prejean} Our
results allow us to infer parameter ranges in which anisotropy and/or
disorder are likely to play an important role in the magnetic
properties of such systems. 

The paper \cite{MRS} is organized as follows: in
Sec.~\ref{sec:model} we describe the models we study and how disorder
and anisotropy are incorporated into each of them.  
Section~\ref{sec:sim} lists the quantities we calculate and how we
perform the Monte Carlo simulations.  
Section~\ref{sec:results} shows our results for the magnetization, susceptibility and Curie
temperature and finally 
in Sec.~\ref{sec:disc} we discuss our results and
their implications for modeling III-V DMS.

\section{Model}
\label{sec:model}
We consider $N_d$ spins randomly distributed at locations
$\bvec{R}_i$, $i = 1, \ldots, N_d$ on a fcc lattice of size $N \times
N \times N$, corresponding to an impurity concentration $x =
N_d/4N^3$.  The spins are treated as classical variables (Mn spins in
Ga$_{1-x}$Mn$_x$As have $S=5/2$, so this is a reasonable approximation). For
simplicity, we take the classical spins to have unit length; this is
equivalent to rescaling the exchange from $J$ to $JS^2$, i.e. changing
the units of energy.  The most general formulation of the problem we
consider in this study is provided by the Heisenberg Hamiltonian
\begin{equation}
{\mathcal H} = - \sum_{i, j}\sum_{\alpha\beta}
J_{ij}F_{\alpha\beta}(\bvec{R}_i - \bvec{R}_j) S_i^\alpha S_j^\beta,
\end{equation}
where $\alpha$ and $\beta$ index Cartesian coordinates. 
The exchange coupling $J_{ij} F_{\alpha\beta}$ is written as 
a product of a random variable $J_{ij}$, and a function $F$ of 
the separation of the two spins, for reasons that will become clear later.
Since the
sites are not on a regular lattice, the summation over site index is
always over all sites of the system. For uncoupled spins $i$ and $j$
we simply have $F_{\alpha\beta}(\bvec{R}_i - \bvec{R}_j)=0$.

\begin{figure}[htb]
\label{cartoon}
\insertpic{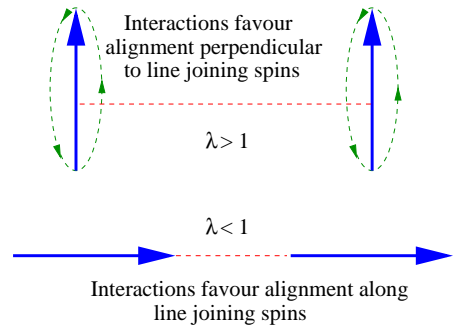}
\caption{ Schematic representation of the two different types of anisotropy parametrized by $\lambda$. 
If $\lambda > 1$, the anisotropy defines an easy plane for the interactions between two spins, whilst
if $\lambda < 1$ the anisotropy defines an easy axis.}
\end{figure}

We consider the exchange integral to be parameterized as:
\begin{equation}
\label{jdef}
F_{\alpha\beta}(\bvec{R}_i - \bvec{R}_j) = \left( \lambda
\delta_{\alpha\beta} + (1-\lambda) \hat{e}_{ij}^\alpha
\hat{e}_{ij}^\beta \right) f(r),
\end{equation}
where $r=|\bvec{R}_i - \bvec{R}_j|$ and the unit vector
$\hat{\bvec{e}}_{ij} = (\bvec{R}_i - \bvec{R}_j)/|\bvec{R}_i -
\bvec{R}_j|$.  
The parameter $\lambda$ controls the exchange anisotropy, 
where $\bvec{e}_{ij}$  defines the axis of anisotropy for the pair of
spins $\bvec{S}_i$ and $\bvec{S}_j$ located at positions $\bvec{R}_i$
and $\bvec{R}_j$.
Using Eq.~(\ref{jdef}), the effective
interaction can now be rewritten as 
$J_{ij}S_i^{\alpha}
F_{\alpha\beta}(\bvec{R}_i - \bvec{R}_j) S_j^\beta =
J_{ij}f(r)\left(S_i^{\parallel} S_j^{\parallel}+ \lambda S_i^{\perp}
S_j^{\perp} \right)$, where the parallel and perpendicular components
are defined with respect to $\hat{\bvec{e}}_{ij}$. As a result, for
$\lambda = 1$ the model has no anisotropy and reduces to a simple
disordered Heisenberg model.  For $0<\lambda < 1$ the couplings become
anisotropic and favor alignment of each pair of spins along their
positional axis $\hat{\bvec{e}}_{ij}$, whilst for $\lambda > 1$, the
couplings favor alignment perpendicular to this axis (see Fig.~\ref{cartoon}). 
Since the
relevant directions $\hat{\bvec{e}}_{ij}$ differ considerably from
pair to pair, some frustration is introduced into the system, possibly
leading to spin-glass physics.

The function $f(r)$ describes the spatial variation of the exchange
interactions.  In this work we investigate the following
possibilities:
\begin{eqnarray}
f(r) = \left\{ \begin{array}{cl} \theta(R_c - r) & 
\mbox{Short range FM, model A,\cite{footnote}} \vspace*{0.2cm} \\
{\displaystyle  \frac{\sin y - y\cos y}{y^4} } & 
\mbox{RKKY, model B,} \vspace*{0.2cm} \\ 
{\displaystyle \frac{1}{r^4}} & 
\mbox{Long range FM, model C,}\vspace*{0.2cm}
\end{array}\right. 
\end{eqnarray}
where $y = 2k_F r$, and $k_F$ is the Fermi wavevector.  In the
short range model, equal-strength ferromagnetic interactions exist only
between neighboring spins within distance $R_c$ of one
another.\cite{footnote} The RKKY model allows interactions 
between all spins in the system, with oscillating sign depending on the 
Fermi wave-vector $k_F$ corresponding to different carrier
 concentrations. 
Finally we consider a purely ferromagnetic model with long-range 
(power-law) interaction.  A comparison
between the short-range and the long-range models clarifies the role
played by the {\it range} of the exchange, while a comparison between
the RKKY and the long-range model clarifies the importance of the
exchange {\it sign oscillations}.

In the RKKY approach of Zarand an Janko,\cite{Zarand} the relative
magnitudes of the exchanges parallel and perpendicular to the line
joining two Mn spins, $K_{par}(r)$ and $K_{perp}(r)$ respectively,
depend on the distance $r$ between the two spins. As a rough guide to
compare with the models we use in this work, $|K_{perp}(r)| >
|K_{par}(r)|$ for large $r$ while $|K_{par}(r)| > |K_{perp}(r)|$ for
small $r$.\cite{Zarand2}
It should be noted that we use the RKKY
interaction for illustrative purposes only, since the exchange
interactions in a more realistic model of DMS will likely be quantitatively different.

As demonstrated in mean-field and Monte Carlo studies of an impurity
band model for III-V diluted magnetic semiconductors,\cite{Zarand2,Twofluid,Mona1,Mona2,MC,JS,CPC,Mayr,Calderon} and in studies of the
kinetic-exchange model of III-V DMS including disorder and Coulomb
interactions of the charge carrier with the Mn acceptors,\cite{Eric}
inhomogeneity induced by positional disorder of the Mn spins 
implies different local carrier charge densities at different sites, 
which in turn
leads to
a broad distribution of effective local fields created at various Mn
sites by the itinerant carriers.  If one integrates out the fermionic
degrees of freedom and formulates the problem in terms of effective
exchanges between the Mn spins, there should also be a wide
distribution of their effective exchanges.  To incorporate the
effect of positional disorder leading to a wide distribution of
magnitudes of the exchanges in our model, we assume that
$J_{ij} = \epsilon_i \epsilon_j$, where the $\epsilon_i$ are random
variables attached to each site, such that their logarithm 
$z_i = \log_{10} \epsilon_i$
has a
Gaussian distribution
\begin{equation}
{\mathcal P}_{n_i}(z_i) = { 1 \over \sqrt{2\pi} \sigma}
\exp\left[ - \frac{(z_i -
z(n_i))^2}{2\sigma^2}\right],
\label{prob}
\end{equation}
although the precise form of the disorder should not change our conclusions.
In Eq.~(\ref{prob}), the mean value of $z_i$, $\bar{z}_i(n_i)$ 
is taken to depend on the number of nearest neighbors spins (i.e. within a distance
$R_c$) of the site $i$,  
$n_i$ though the relation $\bar{z}_i(n_i)= \sigma(n_i-n_0)$,where 
$n_0$ is the average number of neighbors for a given cutoff $R_c$.
 In our simulations $R_c$ is chosen such that $n_0=6$ or $12$.
 
This scheme naturally favors stronger
couplings within more dense clusters (higher $n_i$ values) and hence
should mimic some of the phenomenology observed in the impurity band
model for III-V DMS, where holes congregate in regions of higher Mn
density and therefore lead to stronger effective interactions between
the Mn in these regions.\cite{MC} The parameter $\sigma$
controls the width of the distribution of couplings, and therefore we
use it to characterize the disorder present in the system; a more
disordered system with a wider distribution of couplings corresponds
to a larger value of $\sigma$.

Thus, the three parameters that control the behavior of the model are:
(i) $\lambda$, which controls the amount of anisotropy; (ii) $\sigma$, which
controls the disorder-induced width of the distribution of
effective exchanges, and (iii) $R_c$, which defines the average number of
nearest neighbors for the short range model (for the RKKY and long range models, $R_c=\infty$).

\section{Simulations}
\label{sec:sim}
We have performed exhaustive Monte Carlo simulations on each of models 
A, B and C for temperatures above the ordering temperature $T_c$ to well 
below $T_c$. We use a range of sizes to determine $T_c$ via 
finite size scaling, and average over a sufficient number of 
realizations of the disordered systems for each size to determine average quantities.

\subsection{Parameters and calculated quantities} 
We investigated systems of linear sizes $L = 11$, 14 and 17, 
containing $N_d = 53$, 110 and 196 spins respectively.
This corresponds to a Mn density in Ga$_{1-x}$Mn$_x$As of $x = 0.01$. 
The interaction range $R_c$ is
chosen such that the average number of nearest neighbors within $R_c$
is either $n_0$ = 6, or $n_0 = 12$.  We studied the short range model
for both values of $n_0$.  Note that for models with short-range
interactions, the value of $x$ has no {\it qualitative} effect on the
behavior of magnetic properties for $ x \le 0.05$, since changing
the concentration is equivalent to a pure rescaling of all inter-spin
distances by a fixed factor. The value of $x$ becomes relevant for larger
concentrations, where the average inter-spin distance is comparable
with the lattice constant.  The anisotropy values considered were
in the range $\lambda = 0.1$ to $\lambda = 10$, with the isotropic
case $\lambda = 1$ used as a reference.

\begin{figure}[htb]
\insertpic{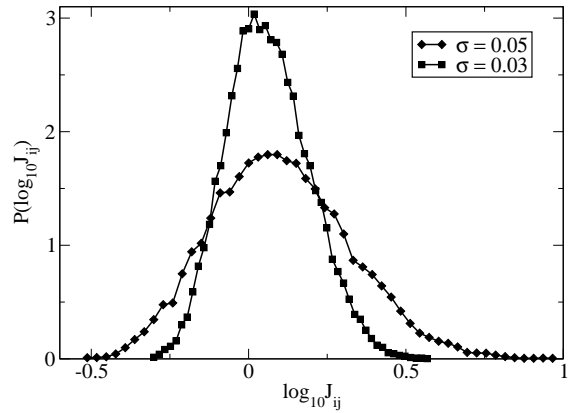}
\caption{ Distribution of $\log_{10}{J_{ij}}$. Each curve is an
average of 5 samples with $L=32$, $N_d = 1310$ and $n_0 = 6$. The
distribution is almost Gaussian, and its width increases with increasing $\sigma$.}
\label{fig:pjij}
\end{figure}

For the short range model, two different disorder values were
considered for most anisotropy values, whilst for the other models,
only one disorder strength was considered.
The values of $\sigma$ considered were: in the short range model,
$\sigma = 0.017$ and 0.05 for $n_0 = 6$,  and $\sigma =
0.01$ and $0.03$ for $n_0 = 12$.  In the long range model we chose
$\sigma = 0.03$, while for the RKKY model, $\sigma = 0.01$. 
These values were chosen such that meaningful comparisons between
different models can be performed, subject to some computational
constraints. Figure~\ref{fig:pjij} shows the actual distribution of $J_{ij}$ used in the simulation for two different values of $\sigma$.

For each pair of $\lambda$ and $\sigma$ values we considered
many realizations of positional disorder of the Mn spins (generally at
least 40) and then averaged over these configurations to obtain our
final results.

We calculated equilibrium (disorder averaged) 
averages for the following quantities: (i) the magnetization
\begin{equation}
M = \left<\frac{1}{N_d}\left|\sum_i \bvec{S}_i\right|\right>,
\end{equation}
and (ii) the magnetic linear susceptibility
\begin{equation}
\chi_m = \beta \left( \left<\frac{1}{N_d}\left|\sum_i
\bvec{S}_i\right|^2\right> - M^2 \right) .
\end{equation}

To determine the Curie temperature in our samples we calculated the Binder
cumulant
\begin{equation}
\label{eq7}
G(L,T) = \frac{1}{2}\left( 5 - 3 \frac{\left<\left|\frac{1}{N_d}\sum_i
\bvec{S}_i\right|^4\right>}{ \left<\left|\frac{1}{N_d}\sum_i
\bvec{S}_i\right|^2\right>^2} \right) ,
\end{equation}
and used finite size scaling. $G(L,T)$ is defined such that in the
paramagnetic phase it decreases with $L$, and tends to zero as $L
\rightarrow \infty$, while in the ferromagnetic phase it increases
with increasing size $L$ and tends to unity in the thermodynamic
limit. Near the transition temperature $T_c$, this dimensionless
quantity has the finite size scaling form $G(L,T) = G[L^{{1 \over
\nu}}(T-T_c)]$ where $\nu$ is the exponent of the diverging spin-spin
correlation length $\xi \sim (T - T_c)^{-\nu}$.\cite{BY,Xin1,Xin2}
Consequently, at $T_c$, $G(L,T_c)$ is independent of $L$; $T_c$ can be
identified by a simultaneous crossing of $G(L,T)$ vs. $T$ curves for
different $L$. This method is found to be more reliable in determining
$T_c$ than the onset of magnetization, or the position of peaks in the
magnetic susceptibility in relatively small finite size samples \cite{MC}.

\subsection{Temperature Rescaling}
In order to compare temperature scales for different $\lambda$ values,
we note that the exchange felt for a spin orientation at angle
$\theta$ to the axis $\bvec{e}_{ij}$ is $J(\theta) =
J_0\sqrt{\cos^2\theta + \lambda^2 \sin^2\theta}$.  Thus,
$\left<J\right> = \frac{1}{4\pi} \int d\Omega \, J(\theta)$ gives the
average exchange integrated over all solid angles
$\mbox{\boldmath{$\Omega$}}$.  Evaluating the integral gives
\cite{GnR}

\begin{eqnarray}
\left< J \right> = J_0 \times \left\{
\begin{array}{cl}
 \frac{1}{2} \left[ 1 + \frac{\lambda^2}{\sqrt{\lambda^2 -1}}
 \sin^{-1}\left(\frac{\sqrt{\lambda^2 - 1}}{\lambda}\right)\right] & ,
 \lambda > 1 , \\ 1 & , \lambda = 1 , \\ \frac{1}{2} \left[ 1 +
 \frac{\lambda^2}{\sqrt{1 - \lambda^2}} \sinh^{-1}\left(\frac{\sqrt{1
 - \lambda^2}}{\lambda}\right) \right] & , \lambda < 1.
\end{array}
\right. \nonumber
\end{eqnarray}
These factors are used to rescale the temperature for each value of
$\lambda$ chosen.  The plots are for the temperature scaled as
$T/J_{\rm eff}(\lambda)$, where $J_{\rm eff}(1) = J_0$, $J_{\rm
eff}(0.1) = 0.5150 \, J_0$, $J_{\rm eff}(0.5) = 0.6901 \, J_0$,
$J_{\rm eff}(1.5) = 1.3463 \, J_0$ and $J_{\rm eff}(10) = 7.8902 \,
J_0$.

\subsection{Monte Carlo Technique}
 
The Metropolis algorithm necessitates long equilibration times at low
temperatures and large values of anisotropy, due to its relatively
slow sampling of phase space.  As a result, we used a different method
to perform spin flips.  This method is formally equivalent to the
Metropolis algorithm and can be used for any Hamiltonian, $\mathcal H$,
 involving
classical spins that can be written in the form ${\mathcal H} = \sum_i
\bvec{h}_i\cdot\bvec{S}_i$, where $\bvec{h}_i = \sum_j J_{ij}
\bvec{S}_j$ is the local field created at site $i$ by the other spins.
In our model, it also contains terms of the form $\sum_{j}{
J_{ij}\bvec{e}_{ij}(\bvec{e}_{ij}\cdot\bvec{S}_j)}$.

The implementation of a Monte Carlo simulation requires successive spin flips 
at each site in the system for a Monte Carlo step.  Each spin flip involves
changing the angular position vector 
of one spin, whilst keeping all others fixed.  After a
sufficiently large number of spin flips at each site, the angular distribution
of these vectors will be equal to the equilibrium (Boltzmann) distribution.
For any spin that is about to be flipped, the
 distribution of the angle
$\theta$ between the spin and its local field $\bvec{h}_i$ (integrated
over azimuthal angle) is
\begin{eqnarray}
 \rho(\theta) = { k e^{k\cos{\theta}} \sin{\theta}\over e^k-e^{-k}},
\end{eqnarray}
where $k =\beta|\bvec{h}_i|$. 
This distribution is easily
integrated over $\theta$, and hence by finding a function $f(x)$ which
maps $x \in [0,1]$ to $\cos\theta \in [-1,1]$, uniform
sampling of the $[0,1]$ interval will give the desired 
distribution of $\cos{\theta} \in [-1,1]$.  It is simple to find 
that $f(x)$ is given by
\begin{eqnarray}
	f^{-1}(\cos\theta) = x = \int_0^\theta \rho(\theta^\prime) d\theta^\prime ,
\end{eqnarray}
and the explicit expression for this mapping is:
\begin{eqnarray}
	\label{eq:master}
	\cos{\theta} = f(x) = 1+ {1 \over k}\ln[1-x(1-e^{-2k})] .
\end{eqnarray} 
This approach is similar to those that are sometimes used in lattice
field theories.\cite{Thijssen} We have tested that our method yields 
results for equilibrium quantities that are identical to the
Metropolis algorithm, at the same time,  the equilibration and auto-correlation times
were found to be shorter by a factor from 10 to 100 depending on 
temperature and size when using this method for the above models for our range of $T$ and $L$.

If the new orientation of the spin $\theta$ is generated in this way,
each trial move is accepted, since the mapping we have made is one that 
changes the probability of a configuration being accepted from the 
Boltzmann weight to unity, provided the configuration is chosen in the
proscribed manner.  The calculation of the mapping
function may be more time-consuming than calculation of the energy difference
between initial and final states performed in the Metropolis algorithm, but
it has the advantage of a 100\% acceptance rate. This compares to the
 Metropolis
algorithm which may have exponentially low acceptance rates. The evaluation of
the mapping function can be optimized to further to increase the
efficiency.  The calculation of $k$ takes about the same amount of
time as evaluating the energy difference in the usual implementation
of the Metropolis algorithm.  After updating $\cos\theta$ all that
remains is to generate a random azimuthal orientation of the
spin by randomly selecting a vector (of appropriate magnitude) 
perpendicular to the local field; this can be done with a fast
algorithm.
 
In summary, the algorithm used is: 1) compute the local field 
  $\bvec{h}_i$ and $k =\beta|\bvec{h}_i|$; 2) choose a random number
  $x\in[0,1]$ and use Eq. (\ref{eq:master}) to generate the new
  orientation $\cos\theta$; 3) update the energy; 4) generate a
  random azimuthal component to the spin; 5) update spin components;
  6) repeat for the next spin.

\begin{figure}
\insertpic{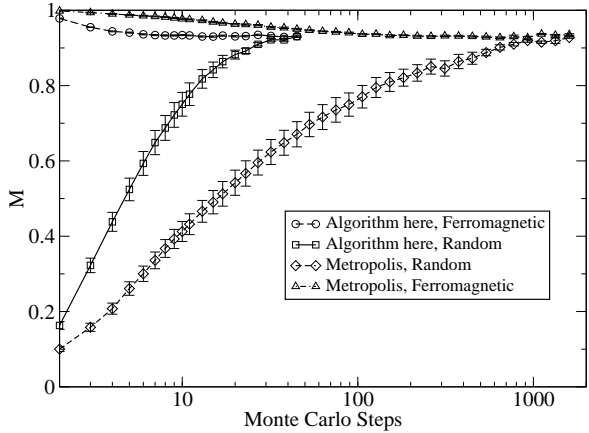}
\caption{A comparison between Metropolis and our algorithm shows the 
convergence to equilibrium of Monte Carlo simulation is accelerated 
about 10 times. This simulation was done with 196 spins, $T=0.5,
\lambda=0.5$ and $\sigma = 0.1$. Each MC step flips 196 spins
randomly. Equilibrium is determined by convergence between replicas 
started with all spins aligned(ferromagnetic initial configuration), 
and random initial configuration. At lower temperatures and with larger disorder, an increase in speed of
more than a factor of 100 was observed.}
\label{fig:acc}
\end{figure}

We have compared this algorithm  with the 
Metropolis algorithm for
our models. They generate the same equilibrium results, however this
algorithm equilibrates much faster when the model has large
anisotropy and frustration. Figure~\ref{fig:acc} shows an example in which
the algorithm we used reached equilibrium an order of magnitude
faster than the Metropolis algorithm.

\section{Results}
\label{sec:results}
We present the results we obtained for the magnetization for all three
models. For the short range model, we also present the susceptibility
and the Curie temperature.

\subsection{Short range model}
\subsubsection{Magnetization}

\begin{figure}[thb]
\insertpic{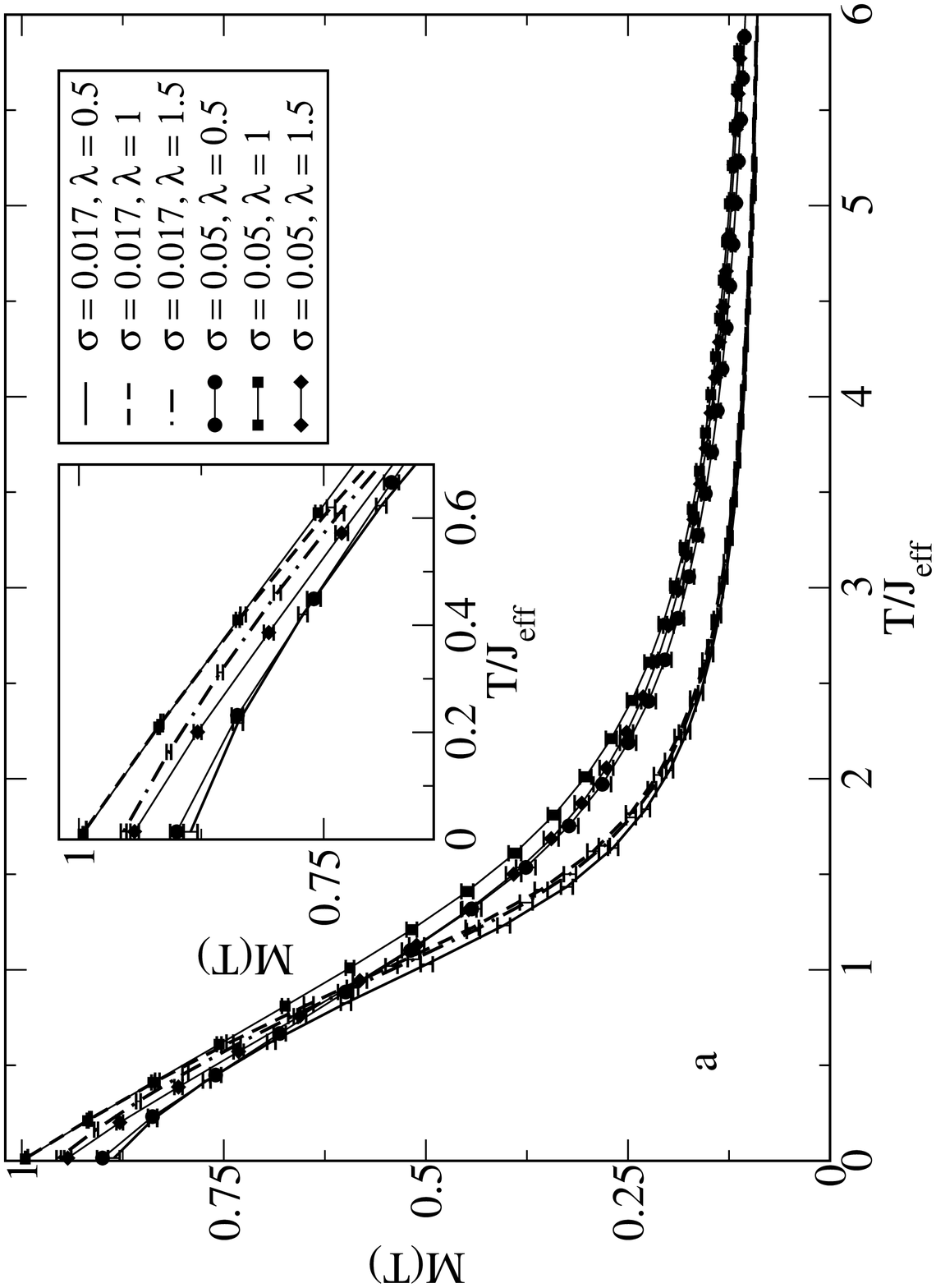} 
\insertpic{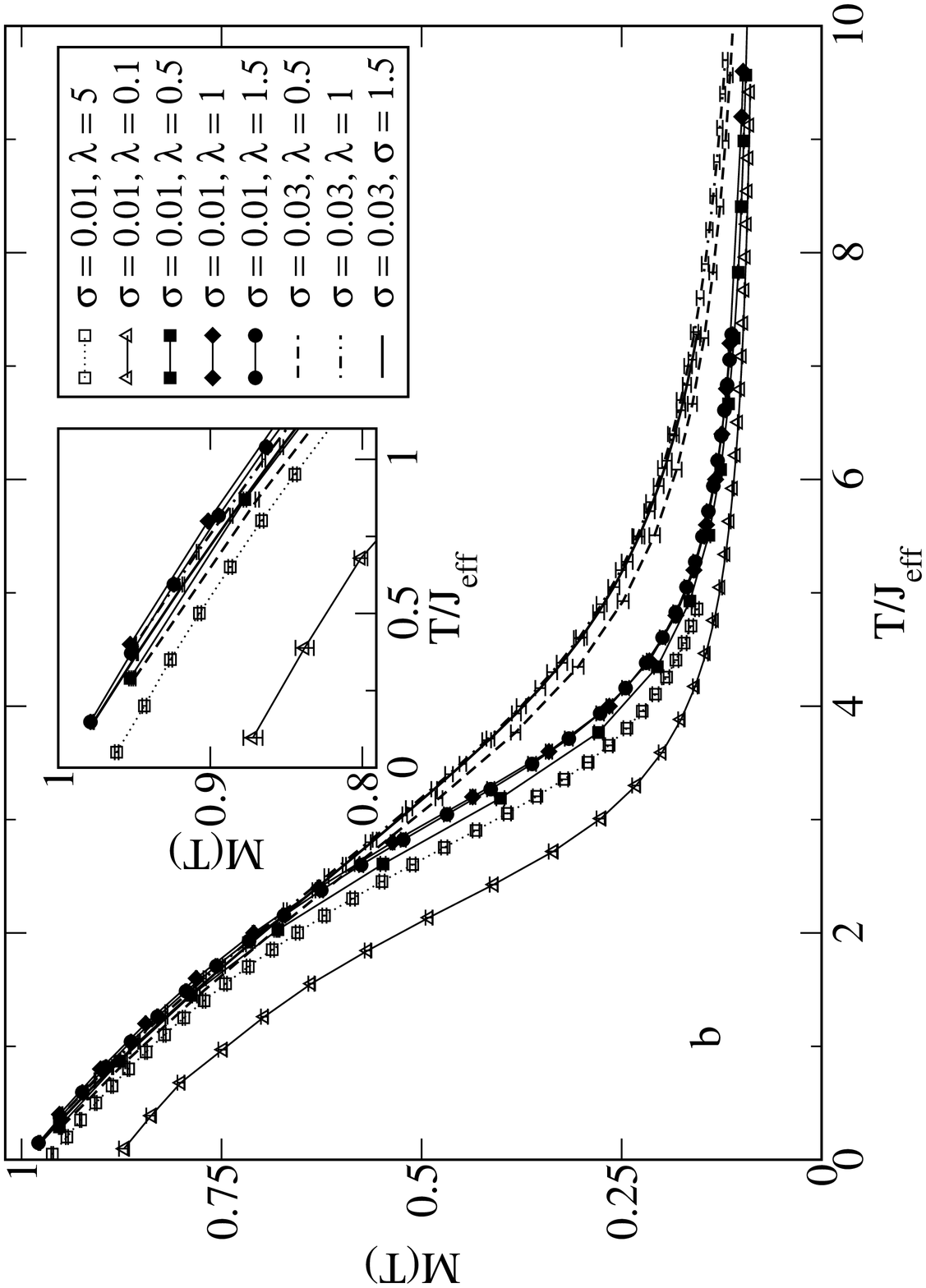}
\caption{Magnetization for (a) $n_0 = 6$ and (b) $n_0=12$, 
for various values of $\lambda$ and $\sigma$. The simulations are for $N_d = 196$ spins. The insets amplify the low temperature regions.}
\label{fig:magnshort}
\end{figure}

The magnetization curves $M(T)$ shown in Fig.~\ref{fig:magnshort} for size corresponding to $N_d= 196$ spins have
the characteristic linear decrease with temperature seen in previous
work and in experiments in Ga$_{1-x}$Mn$_x$As.\cite{Beschoten,Ohno} It is also
 apparent from comparing Fig.~\ref{fig:magnshort}(a)  with Fig.~\ref{fig:magnshort}(b) that increasing the
number of nearest neighbors in the short range model does not qualitatively change
the {\em shape} of the magnetization curve.

With temperatures properly rescaled by $J_{\rm eff}$, we observed
that curves corresponding to the same disorder value $\sigma$ but
various anisotropies $\lambda$ are almost identical at high
temperatures. Increasing the disorder $\sigma$ leads to increased
magnetization at high temperatures, as expected from studies of other
models for DMS.\cite{Mona1,Mona2,MC}

Anisotropy plays a role at low temperatures, where increased
anisotropy does lower the $T=0$ magnetization, although the
effect is rather small. The suppression is more pronounced for
$\lambda <1$ (easy-axis) than $\lambda > 1$ (easy-plane).  For
$n_0=6$, the value of the saturation magnetization seems to depend
only on $\lambda$ and be independent of $\sigma$.  
Whereas at higher  temperatures ($T\ge 0.5 \, T_c$), as 
stated above the opposite is true. Consequently, 
both anisotropy and disorder in the magnitude of the 
ferromagnetic exchanges affect $M(T)$ in 
the temperature range $T = 0.1 \, T_c \sim 0.5 \, T_c$. For $n_0 = 12$, the
suppression of low-temperature magnetization by anisotropy is not
observed except for an extremely large anisotropy $\lambda=0.1$.

\subsubsection{Susceptibility}
\begin{figure}[htb]
\insertpic{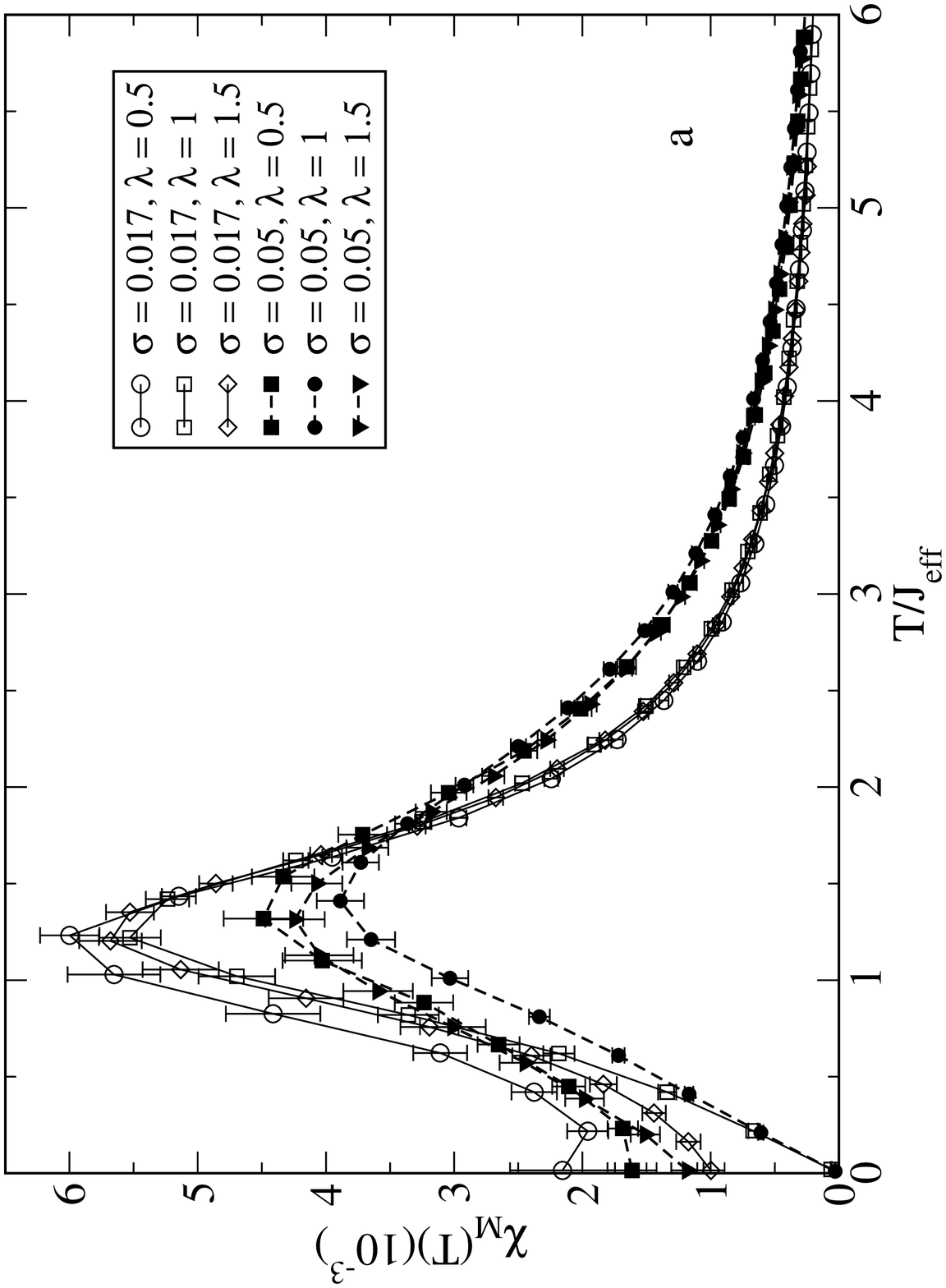} 
\insertpic{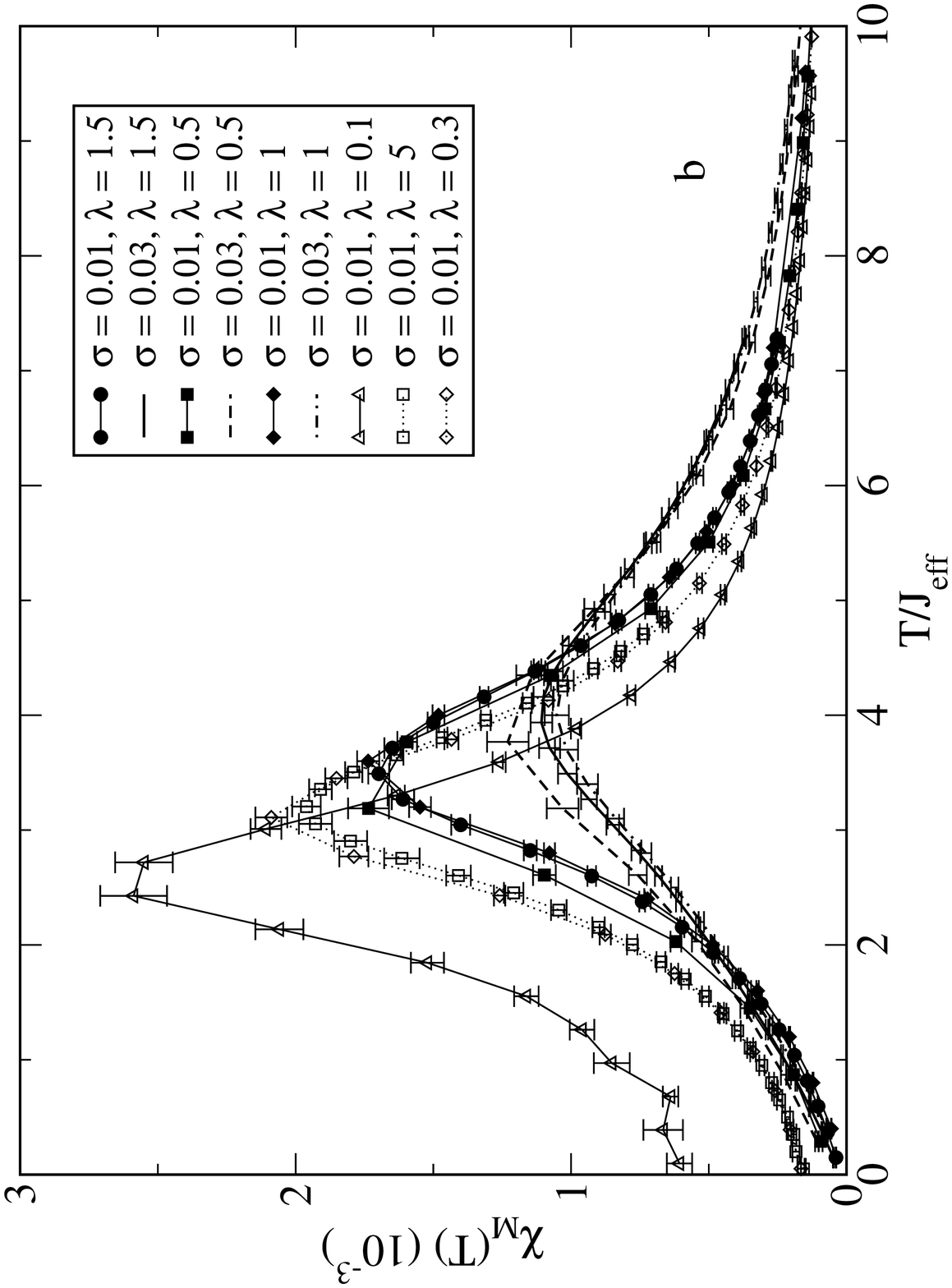}
\caption{ Magnetic  susceptibility for (a) $n_0 = 12$,
and (b) $n_0 = 6$, with various values of $\lambda$ and $\sigma$. The simulations are for  $N_d = 196$ spins.}
\label{fig:suscn}
\end{figure}

The behavior of the linear susceptibility for
the short range model with both 6 and 12 nearest neighbors is shown
in Fig.~\ref{fig:suscn}.  After temperature
rescaling, the high-temperature tails of the curves with the
same $\sigma$ value are again identical. At low
temperatures, anisotropy leads to a finite value for the $T=0$
linear susceptibility.
The finite value of the linear susceptibility at $T=0$ is consistent
with the incomplete saturation of magnetization in the presence of
anisotropy. This effect is most
transparent in the case of $n_0=6$; in contrast, for $n_0=12$,
the effect of anisotropy is very weak unless the anisotropy is
extreme, e.g. $\lambda>5$ or $\lambda<0.1$.

\subsubsection{Curie Temperature}

\begin{figure}[htb]
\insertpic{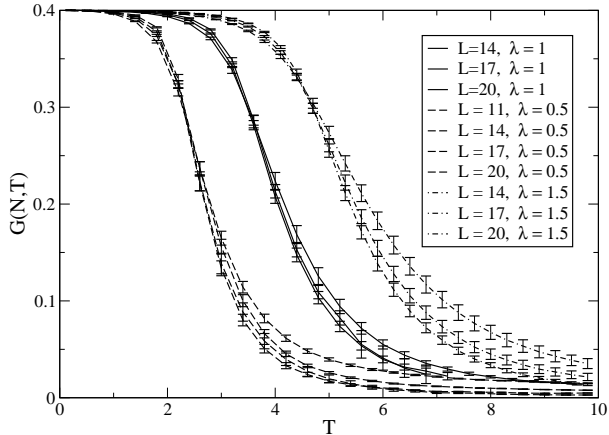}
\caption{Binder cumulant for models with 12 neighbors on average,
	$\sigma = 0.01$, various anisotropy $\lambda$ and sizes $L$.}
\label{fig:Curie}
\end{figure}

The Curie temperatures are deduced from the Binder cumulant Eq. (\ref{eq7})
curves. Figure~\ref{fig:Curie} shows curves calculated for different
system sizes from the 12 neighbor model which intersect at the same
temperature for a given $\lambda$, indicating the position of $T_c$. These
critical temperatures are very close to the peaks in linear
susceptibility of the same model, and are tabulated in Table \ref{tab1}. Note that the temperature is not rescaled by
$J_{\rm eff}$ in Fig.~\ref{fig:Curie} in order to give a clear view of each
set of curves. When the temperature is rescaled, the values of $T_c$ are
approximately the same.

\begin{table}[hbtp]
\begin{ruledtabular}
\begin{tabular}{l|c c c} 
 $ T_c / J_{eff}$ & $\lambda = 0.5$ & $\lambda = 1 $ & $\lambda = 1.5$ \\ \hline 
 From $\chi_M$    & 3.42            & 3.61           & 3.50   \\ \hline
 From $G(N,T)$    & 3.39            & 3.48           & 3.39   \\ 
\end{tabular}
\end{ruledtabular}
\caption{Critical temperatures (in units of $J_{eff}$, see text) estimated from Binder cumulant $G(N,T)$ and magnetic susceptibility $\chi_M$ for $\sigma = 0.01$ and $n_0 = 12$.}
\label{tab1}
\end{table}

We conclude that small-to-moderate anisotropy has a rather small
effect on the magnetic properties of the short-range
model. Significant deviations from the isotropic behavior are seen only
for very large values of the anisotropy, and are generally more
pronounced in the $n_0 = 6$ model.

\subsection{Long range model}
\begin{figure}[htb]
\insertpic{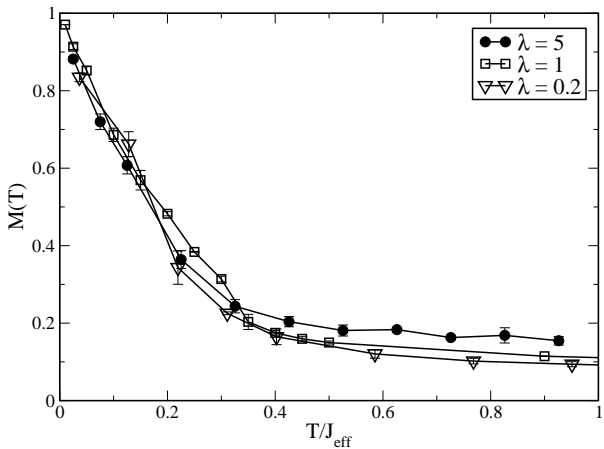}
\caption{Magnetization for $n_0 = 6$, various $\lambda$, $\sigma = 0.03$,
long range model. In this case all pairs of spins interact, and
$n_0$ is used only to adjust the average coupling constant.} 
\label{fig:maglr}
\end{figure}

The qualitative behavior of the low-temperature magnetization in
the long range model is very similar to that observed in the
short range model, as can be seen in Fig.~\ref{fig:maglr}. Increasing
the length-scale of the exchange does very little to increase the
amount of frustration in the model, and in both the short-range
and the low-range model the suppression of the low temperature
magnetization is less than 20\%, and often much smaller, for all values of anisotropy
considered. This is small compared to the 50-60\% reduction observed
experimentally,\cite{Oiwa} hence other sources of frustration must be
present to account for the greatly reduced low temperature
magnetization as compared to that of an aligned ferromagnet. 
One candidate for the source of further reduction is the
antiferromagnetic component of effective Mn-Mn interactions as is found
for RKKY interactions.

\subsection{RKKY models}

\begin{figure}[t]
\insertpic{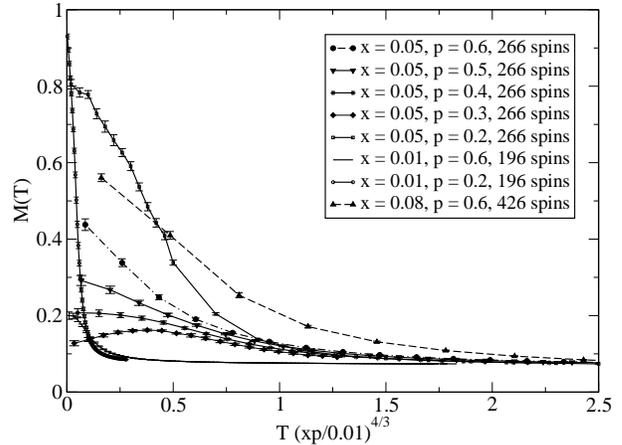}
\caption{Magnetization for the isotropic, $\lambda = 1$ RKKY model, at
  various values of the Mn concentration $x$ and hole compensation
  $1-p$ (corresponding to different choices of $y = 2k_F r$). }
\label{fig:RKKYxp}
\end{figure}

Figure \ref{fig:RKKYxp} shows the magnetization calculated for
isotropic RKKY models with different Fermi wavelengths,
corresponding to different charge carrier concentrations. The main
observation is that the antiferromagnetic exchange at long distances
introduced by the RKKY model has a significant effect on
lowering the low temperature magnetization; the decrease is
considerably larger than that due to anisotropy in purely
ferromagnetic exchange interactions in the short- and long-range
models considered previously.

  In Fig.~\ref{fig:RKKYxp} we show curves corresponding to several
  different values of the Mn concentration $x$ and charge carrier
  compensation $1-p$; the resulting charge carrier concentration is
  $n_h = 4xp/a^3$, where $a=5.65$ \AA \, is the GaAs lattice constant.
  Note that the temperature also needs to be rescaled by a factor of
  $(px)^{\frac{4}{3}}$ in 3 dimensions for RKKY interactions,
  \cite{Dietl} to compare different magnetization curves.  One
  interesting observation of the series with $x=0.05$ is that
  increasing the hole concentration $px$ at with fixed $x$ does not
  appear to change $T_c$ but the low temperature magnetization has a minimum
   at about $p = 0.3$. Further
  increments of $p$ increase the low temperature magnetization
  gradually.  There is a large range of values of the $T=0$
  magnetization that can be achieved by tuning the Mn and the hole
  concentrations. This can be understood qualitatively in the
  following way: larger hole concentrations $px$ lead to larger Fermi
  wavevectors $k_F$. As a result, the oscillation between
  ferromagnetic and antiferromagnetic interactions described by the
  RKKY interaction appears at a shorter distance. If this distance
  becomes comparable to the average inter-spin distance, significant
  frustration is present in the system leading to suppressed low-T
  magnetization. 
  When $p=0.3$, the antiferromagnetic part seems to be dominant. The
  magnetization actually decreases as temperature is lowered below
  0.4$\,J_{\rm eff}$. This effect
  arises due to antiferromagnetic coupling between clusters as a result
  of the long-range nature of the RKKY interaction.\cite{Priour}

  We have also
  investigated RKKY models with exchange anisotropy as discussed
  before. In all cases, the additional suppression of the low-T
  magnetization induced by the anisotropy is very small (less than
  10\%).

\section{Discussion}
\label{sec:disc}
The major result that can be deduced from our simulation is that while exchange anisotropy
can change the temperature scale at which ferromagnetism occurs in a
model, it does not change the essential {\em qualitative} features
such as the shape of the magnetization curve, or the form of the
linear or susceptibility greatly above $T_c$, unless the ratio between
the parallel and perpendicular exchanges is extremely different from
1. Below $T_c$, the anisotropy preserves a finite magnetization and
linear susceptibility at $T=0$.
This effect is suppressed by higher connectivity (large $n_0$). 
[Locally the random lattice with $n_0=12$ is more symmetric than
 for $n_0 = 6$, so that the anisotropy a spin feels from one 
neighbor is more likely to be cancelled by the effects of other
neighbors.]
 However, in general the effect of
anisotropy is relatively weak, as it does not lower magnetization at $T=0$ by
more than 20\%.  
The presence of antiferromagnetic interactions, as occur in 
oscillatory exchange interactions, such as in the RKKY model, appears to be
much more important for lowering the magnetization at low temperature.
Thus the large reduction in the magnetization seen in Ref.~\onlinecite{Zarand} can
probably be attributed to RKKY interactions, rather than the effects of anisotropy
in the exchanges.

Our result on the effects of moderately 
anisotropic exchange interactions on the 
magnetic properties of DMS are in substantial contrast with the results of Zarand 
and Janko.\cite{Zarand}  They found large differences between the low
temperature magnetization (but similar Curie temperatures) in the isotropic
and anisotropic cases.  We are puzzled by this difference, but suggest that 
it may be due to the fact that they used an exponential cut-off in
their exchange interaction.  This cut-off may have damped the effect of
anti-ferromagnetic interactions in the isotropic case, but since the parallel 
and perpendicular exchanges had different spatial dependences in their study, 
they may not have reduced the effect of anti-ferromagnetic exchanges in the anisotropic case.

There are a number of other interesting observations.  
The two types of anisotropy, corresponding to the cases
$\lambda < 1$ and $\lambda > 1$ are qualitatively different.  For
$\lambda < 1$, the spins preferentially align along the line joining
them, whilst for $\lambda > 1$ there is a preferred plane in which the
spins may lie.  This difference in dimensionality appears to explain
why the magnetization is not suppressed as much at low temperatures
for $\lambda > 1$ as compared to $\lambda < 1$, since it is easier for 
spins to relax from frustration in two rather than one dimensions.

Whilst anisotropic exchange interactions are one possibility which can 
lead to a reduction in the saturation magnetization at low temperatures, 
there are a number of other possibilities.  Firstly, there are believed to be 
antiferromagnetic  nearest neighbor interactions between Mn spins, 
that should have little effect due
to the diluteness of the Mn spins,
but could lead to a decrease in the saturation magnetization.\cite{DasSarma} 
Next is a theoretical proposal
that there is an instability purely in the presence of disorder towards a 
non-collinear ground state.\cite{Schliemann,Chudnovskiy}  Another is the
experimental observation that there are significant numbers of interstitial 
Mn that appear to be involved in compensation processes and thus do not polarize.\cite{Yu}
It is not completely clear to what extent the saturation at low temperatures in
Ga$_{1-x}$Mn$_x$As is due to non-participation of Mn spins
{\em due to the presence of Mn interstitial defects.} 
However, given the recent progress in this area,\cite{Yu}
it will probably not be long before an accurate estimate of the proportion of 
Mn that are participating in the ferromagnetism is known. When this is 
quantified, it should be possible to determine whether anisotropy need be included in 
realistic models of DMS and if so, how much.  The carrier-mediated
nature of the ferromagnetism also suggests that anisotropy may  
be important for transport, especially in the insulating phase, since
hopping between sites will be preferred when the Mn spins have similar 
orientations.\cite{Burkov}

We mention in passing that while anisotropy has small effect on the magnetization, it seems to have significant effect on the non-linear susceptibility at low temperature. \cite{unpublished} This
suggests that experimental 
investigations of the non-linear susceptibility might shed light on the magnetic state 
in DMS.  It was recently suggested that for $x < 0.01$ and $x>0.1$ in Ga$_{1-x}$Mn$_x$As,
there may be a spin glass phase.\cite{DasSarma} Here we suggest that if there are
strongly anisotropic exchange interactions, signatures might be seen in co-existence with
ferromagnetism.  If such spin-glass signatures are not seen in (Ga,Mn)As, they may be
present in other insulating materials with lower carrier concentrations such as Ge:Mn.\cite{Kim}

In conclusion, there are still a number of outstanding questions as to the
nature of the ferromagnetic state in DMS, and whether anisotropy plays an
important role in these materials.  This work should be of help in clarifying
which types of models are likely to be affected by anisotropy and what types of experimental probes might help to detect it.

\section{Acknowledgements}
The authors would like to thank Gergely Zarand for stimulating	
conversations, and the Princeton Computer Science department for	
access to some of their computer power.  This research was supported
by NSF grant DMR-9809483 and 0213706(CZ, MPK, MB, and RNB) and DMR-9971541(XW). 
X.W. acknowledges support from the State of Florida. 
M.B. acknowledges support from the Natural Sciences and Engineering Research Council of Canada.
X.W., M.B. and R.N.B also thank the Aspen Institute for Physics for hospitality while
parts of this work were carried out.


\begin{references}
\bibitem{footnote} To be more precise, each spin is coupled to at
least two other neighboring spins.  If a spin has less than 2
neighbors within a distance $R_c$, then we assume it to be coupled to
its two closest neighbors (irrespective of how far they are).  This
avoids the existence of completely isolated spins in our model. For
the values of $R_c$ we investigate, we find that only about 1-2\% of
spins are in this situation.
\end{references}
\end{document}